\documentclass{iau}
\usepackage{graphicx}

\title[On the uniqueness of kinematical signatures of IMBHs in globular clusters]
{On the uniqueness of \\ kinematical signatures of intermediate-mass
black holes in globular clusters}

\author[Alice Zocchi, Mark Gieles, \& Vincent H\'{e}nault-Brunet]   {Alice Zocchi$^1$, Mark Gieles$^2$, \and Vincent H\'{e}nault-Brunet$^3$}

\affiliation{Department of Physics, University of Surrey, \\ Guildford, Surrey, GU2 7XH, United Kingdom \\[\affilskip]
$^1$email: {\tt a.zocchi@surrey.ac.uk}\\
$^2$email: {\tt m.gieles@surrey.ac.uk}\\
$^3$email: {\tt v.henault-brunet@surrey.ac.uk}} 

\pubyear{2015}
\volume{312}  
\pagerange{yyy--yyy}
\jname{Star Clusters and Black Holes in Galaxies across Cosmic Time}
\editors{}
\begin{document}

\maketitle

\begin{abstract}
Finding an intermediate-mass black hole (IMBH) in a globular cluster (GC), or proving its absence, is a crucial ingredient in our understanding of galaxy formation and evolution. The challenge is to identify a unique signature of an IMBH that cannot be accounted for by other processes. Observational claims of IMBH detection are often based on analyses of the kinematics of stars, such as a rise in the velocity dispersion profile towards the centre. In this contribution we discuss the degeneracy between this IMBH signal and pressure anisotropy in the GC. We show that that by considering anisotropic models it is possible to partially explain the innermost shape of the projected velocity dispersion profile, even though models that do not account for an IMBH do not exhibit a cusp in the centre.
\end{abstract}

\firstsection 

\section{Introduction}

The task of finding an intermediate-mass black hole in a globular cluster is very challenging, because the predicted signatures of an IMBH are degenerate with alternative scenarios. Intermediate-mass black hole detections in globular clusters are mostly claimed on the basis of the discovery of a shallow cusp in the surface brightness profile and a rise in the velocity dispersion profile towards the centre (e.g., see \cite[Noyola et al. 2008]{Noyola2008}, \cite[van der Marel \& Anderson 2010]{vdMA2010}, \cite[Noyola et al. 2010]{Noyola2010}). However, similar features can also be produced by different processes, and conclusive evidence for the existence of IMBHs in globular clusters is still lacking. For example, mass segregation, core collapse, or the presence of binary stars in the centre can also generate a shallow cusp in the surface brightness profile, as shown by means of dedicated N-body simulations (\cite[Vesperini \& Trenti 2010]{VT2010}). Some kinematical properties can be explained by the presence of pressure anisotropy, without a central intermediate-mass black hole (\cite[van der Marel \& Anderson 2010]{vdMA2010}; \cite[Zocchi et al. 2012]{ZBV2012}). In order to find conclusive evidence of the presence of IMBH in globular clusters, it is important to fully understand the internal dynamics of these systems: in this contribution we focus on the role played by radially-biased pressure anisotropy.

\begin{figure}[t]
\begin{center}
 \includegraphics[width=0.84\textwidth]{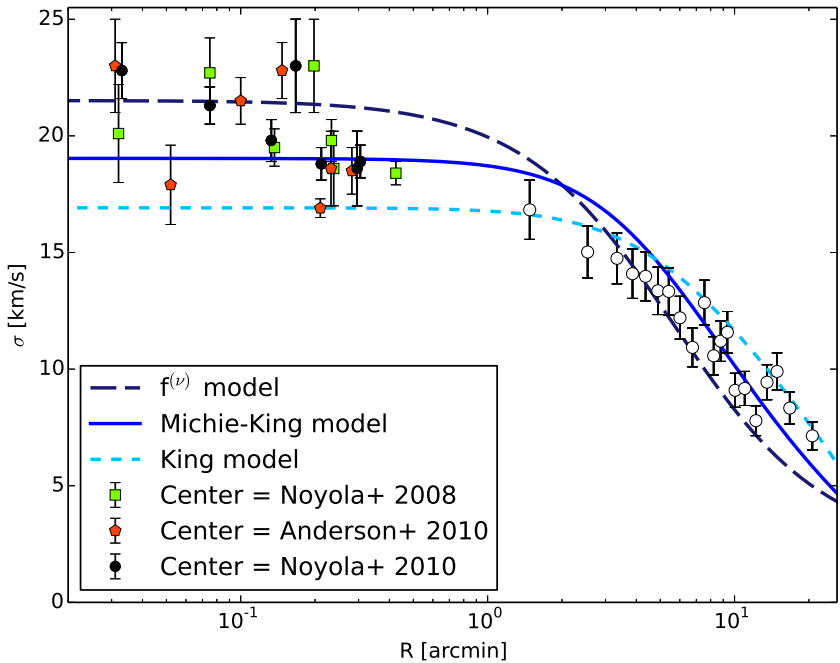} 
 \caption{Velocity dispersion profile of $\omega$ Cen. Best-fit $f^{(\nu)}$ (long-dashed line), Michie-King (solid line), and King (short-dashed line) models profiles are shown along with line-of-sight velocity dispersion data. Empty circles represent the velocity dispersion profile calculated from velocities of single stars as in \cite{BVBZ13}. Filled symbols represent the velocity dispersion calculated from integrated spectra observations carried out by \cite{Noyola2010}, and binned with respect to the position of the centre proposed by \cite[Noyola et al. 2008]{Noyola2008} (squares), \cite[van der Marel \& Anderson 2010]{vdMA2010} (pentagons), and \cite[Noyola et al. 2010]{Noyola2010} (circles); the shape of these profiles is highly dependent on the position of the center, and only in one case a cusp is visible.}
   \label{VD}
\end{center}
\end{figure}
\begin{figure}[t]
\begin{center}
 \includegraphics[width=0.49\textwidth]{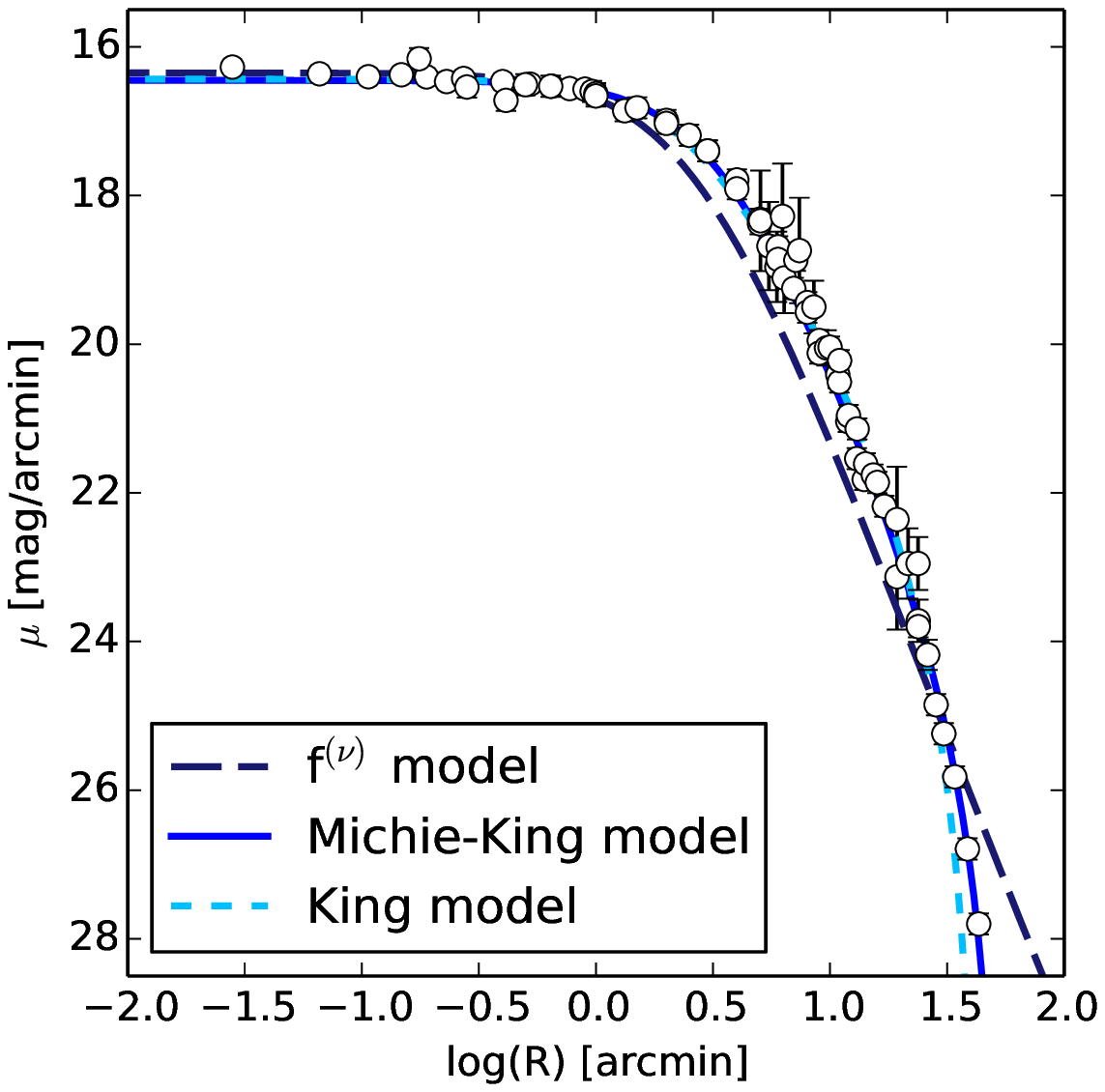} 
 \includegraphics[width=0.49\textwidth]{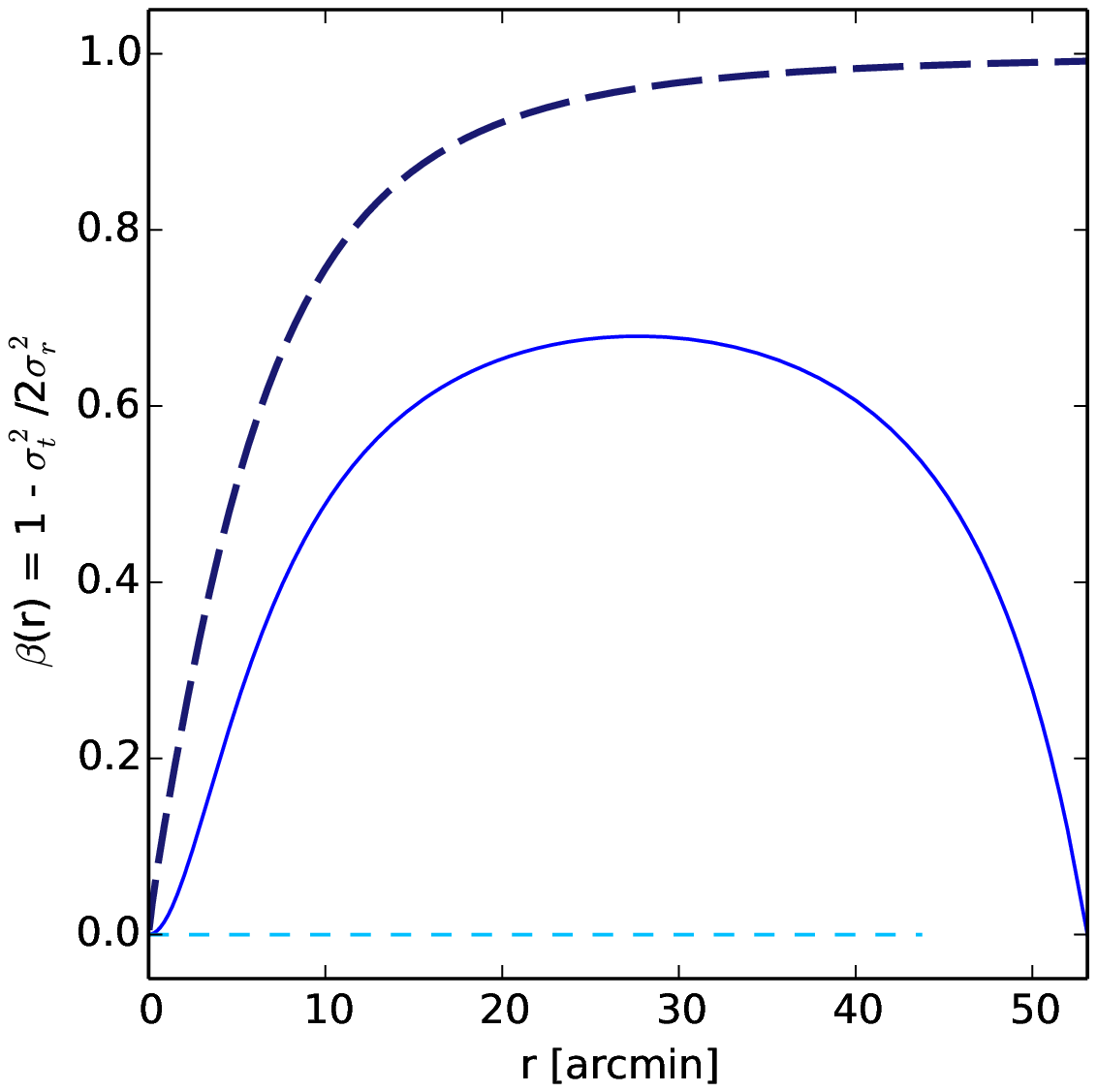} 
 \caption{Left panel: best-fit surface brightness profiles and data from \cite{TKD1995} and \cite{Noyola2008}. Right panel: best-fit models anisotropy profiles. Lines are as in Fig.~\ref{VD}.}
   \label{SB-beta}
\end{center}
\end{figure}

\section{Why is pressure anisotropy relevant?}

Galactic globular clusters are characterized by different relaxation conditions. For many of them, the relevant relaxation times are shorter than their age, so that they are commonly considered to be close to thermodynamical equilibrium, with an isotropic velocity distribution that is close to Maxwellian. However, some large globular clusters have very long relaxation times, and their structure might be more similar to that of elliptical galaxies, for which pressure anisotropy is thought to play an important role.

Several studies have suggested the presence of pressure anisotropy in globular clusters (\cite[Ibata et al. 2011]{Ibata2011}, \cite[Zocchi et al. 2012]{ZBV2012}, \cite[Bellazzini et al. 2015]{Bellazzini2015}), by analysing sets of line-of-sight kinematical data. The only data that would enable us to directly measure the presence of anisotropy in these stellar systems are proper motions (see for example \cite[Bianchini et al. 2013]{BVBZ13}), but unfortunately these are available only for a few GCs\footnote{Measurement of proper motions for several Galactic globular clusters are becoming available via the HSTPROMO collaboration \cite[(Bellini et al. 2014)]{Bellini2014}, and even more will be available in the near future thanks to the observations carried out by the Gaia satellite.}.

Confirmations of the importance of this physical ingredient arrive also from numerical simulations. \cite{Luetz2011} showed that primordial pressure anisotropy can last for long in the outer parts of globular clusters, while in the central region it is not stable, and is washed away very quickly. We recently found that anisotropy can originate in the early phases of the life of GCs, even for systems that are originally isotropic, as the result of two-body relaxation that scatters stars out of the core on radial orbits; for clusters located in an external tidal field, it is erased in the final stages, just before their complete dissolution (Zocchi et al., in prep). 

Even if pressure anisotropy might not concern the very central part of GCs, where we look for the presence of an IMBH, we need to determine its role in shaping the \textit{projected} kinematical quantities that we measure, especially when using only line-of-sight kinematical data. Indeed, the presence of radially-biased anisotropy has the effect of increasing the velocity dispersion that is measured when looking towards the centre of the system, where the line-of-sight is aligned with the radial direction; similarly, in the outermost parts, where the line-of-sight is parallel to the tangential direction, the measured velocity dispersion would be smaller (the opposite is true for tangentially-biased anisotropic systems). It is therefore crucial to take anisotropy into account when looking for the presence of an intermediate-mass black hole in the centre of globular clusters.

\section{The case of $\omega$ Cen}

To illustrate the effect that anisotropy has on projected velocity dispersion profiles, we analyse the data available for the massive globular cluster $\omega$ Cen (NGC 5139), for which several studies have been carried out, looking for the presence of an IMBH at its centre (\cite[Noyola et al. 2008]{Noyola2008}, \cite[van der Marel \& Anderson 2010]{vdMA2010}, \cite[Noyola et al. 2010]{Noyola2010}). 

In this contribution we consider three different families of dynamical models; for simplicity, and to show more clearly the contribution of anisotropy, we chose to consider only single-mass, spherical, non-rotating models. The nontruncated radially-biased anisotropic \textbf{$\mathbf{f^{(\nu)}}$ models} (see \cite[Bertin \& Trenti 2003]{BertinTrenti2003}, and references therein) were constructed to describe the products of (incompletely) violently relaxed elliptical galaxies, and describe systems that are isotropic in the centre, and anisotropic in the outer parts; the anisotropy profile has roughly the same shape for all the models in the family. Truncated radially-biased anisotropic \textbf{Michie-King models} (\cite[Michie 1963]{Michie1963}) describe systems that are isotropic in the centre and near the tidal radius, and anisotropic in the intermediate radial range. Depending on the value of a model parameter, namely the anisotropy radius $r_{a}$, it is possible to obtain different shapes for the anisotropy profile, with different values for the maximum of the anisotropy $\beta$: when $r_{a}$ is very small, the maximum of $\beta$ tends to 1, while for very large values of $r_{a}$ the profile becomes the same as for an isotropic model. For comparison, we also consider isotropic \textbf{King models} (\cite[King 1966]{King66}).

We performed fits of these models to the surface brightness and velocity dispersion profiles of $\omega$ Cen. Figure~\ref{VD} shows the best-fit velocity dispersion profiles along with line-of-sight data (see caption for a description). The effect of anisotropy is clearly visible: the central velocity dispersion is larger for the anisotropic $f^{(\nu)}$ models, and smaller for King isotropic models. By considering anisotropic models it is therefore possible to partially explain the innermost shape of the velocity dispersion profile, even though models that do not account for an IMBH exhibit a flat profile and no cusp in the centre. Anyway, it is clear that, by taking anisotropy into account, a smaller IMBH mass would be needed to match a central cusp. The left panel of Fig.~\ref{SB-beta} shows the best-fit surface brightness profiles: the models represent reasonably well the data; only the $f^{(\nu)}$ model shows some significant deviations from the observed profile. The right panel of Fig.~\ref{SB-beta} shows the anisotropy profiles of the best-fit models; we checked that the anisotropic models are stable against radial orbit instability, by computing the value of the parameter $\kappa$ (1.75 and 1.36 for $f^{(\nu)}$ and Michie-King models, respectively) introduced by \cite{PS1981}. This amount of energy in radial orbits is consistent with the build-up of radial orbits that we find in our N-body models (Zocchi et al. in prep).

\section{Conclusion}

Pressure anisotropy plays an important role in the dynamics of globular clusters, and it should be taken into account to properly describe these systems. Here we showed that models with isotropic velocity distributions in the core and radial anisotropy in the outer parts can describe reasonably well the surface brightness and velocity dispersion profiles of $\omega$ Cen, without the need of the presence of an IMBH. The models we used do not take into account some of the known complexities of globular clusters: by considering rotation and by including a range of stellar masses, it will be possible to give a more accurate representation of the dynamics of this system.

The uncertainty in the position of the centre of the cluster prevents an accurate determination of kinematic and photometric profiles. In the future, we plan to adopt a discrete fitting approach, and we will determine the position of the centre as a fitting parameter.

\end{document}